\documentclass[sn-nature]{sn-jnl}

\usepackage{graphicx}%
\usepackage{multirow}%
\usepackage{amsmath,amssymb,amsfonts}%
\usepackage{amsthm}%
\usepackage{mathrsfs}%
\usepackage[title]{appendix}%
\usepackage{xcolor}%
\usepackage{textcomp}%
\usepackage{manyfoot}%
\usepackage{booktabs}%
\usepackage{algorithm}%
\usepackage{algorithmicx}%
\usepackage{algpseudocode}%
\usepackage{listings}%

\usepackage[T1]{fontenc}
\usepackage{lmodern}
\usepackage[utf8]{inputenc}
\usepackage[english]{babel}

\usepackage{xcolor}

\begin{document}

\title{Non-Hermitian polariton-photon coupling in a perovskite open microcavity}

\author[1]{M. Kędziora}
\author[1,2]{M. Król}
\author[1]{P. Kapuściński}
\author[1,4]{H. Sigurðsson}
\author[3]{R. Mazur}
\author[3]{W. Piecek}
\author[1]{J.Szczytko}
\author[5,6]{M. Matuszewski} 
\author[1,5]{A. Opala} 
\author*[1]{B. Piętka}\email{Barbara.Pietka@fuw.edu.pl}

\affil[1]{Institute of Experimental Physics, Faculty of Physics, University of Warsaw, ul. Pasteura 5, PL-02-093 Warsaw, Poland}

\affil[2]{ARC
Centre of Excellence in Future Low-Energy Electronics Technologies and
Department of Quantum Science and Technology and Research School of Physics, The Australian National University, Canberra, ACT, 2601, Australia}

\affil[3]{Science Institute, University of Iceland, Dunhagi 3, IS-107, Reykjavik, Iceland}

\affil[4]{Institute of Applied Physics, Military University of Technology, Warsaw, Poland}

\affil[5]{Institute of Physics, Polish Academy of Sciences, Aleja Lotników 32/46, 02-668 Warsaw, Poland}

\affil[6]{Center for Theoretical Physics, Polish Academy of Sciences Aleja Lotników 32/46, 02-668 Warsaw, Poland}

\abstract{
Exploring the non-Hermitian properties of semiconductor materials for optical applications is at the forefront of photonic research. However, the selection of appropriate systems to implement such photonic devices remains a topic of debate. In this work, we demonstrate that a perovskite crystal, characterized by its easy and low-cost manufacturing, when placed between two distributed Bragg reflectors with an air gap, can form a natural double microcavity. This construction shows promising properties for the realisation of novel, tunable non-Hermitian photonic devices through strong light-matter coupling. We reveal that such a system exhibits double-coupled polariton modes with dispersion including multiple inflection points. Owing to its non-Hermiticity, our system exhibits nonreciprocal properties and allows for the observation of exceptional points. Our experimental studies are in agreement with the theoretical analysis based on coupled mode theory and 
calculations based on transfer matrix method.}





\maketitle
\begin{figure*}[b]
\includegraphics[width=\textwidth]{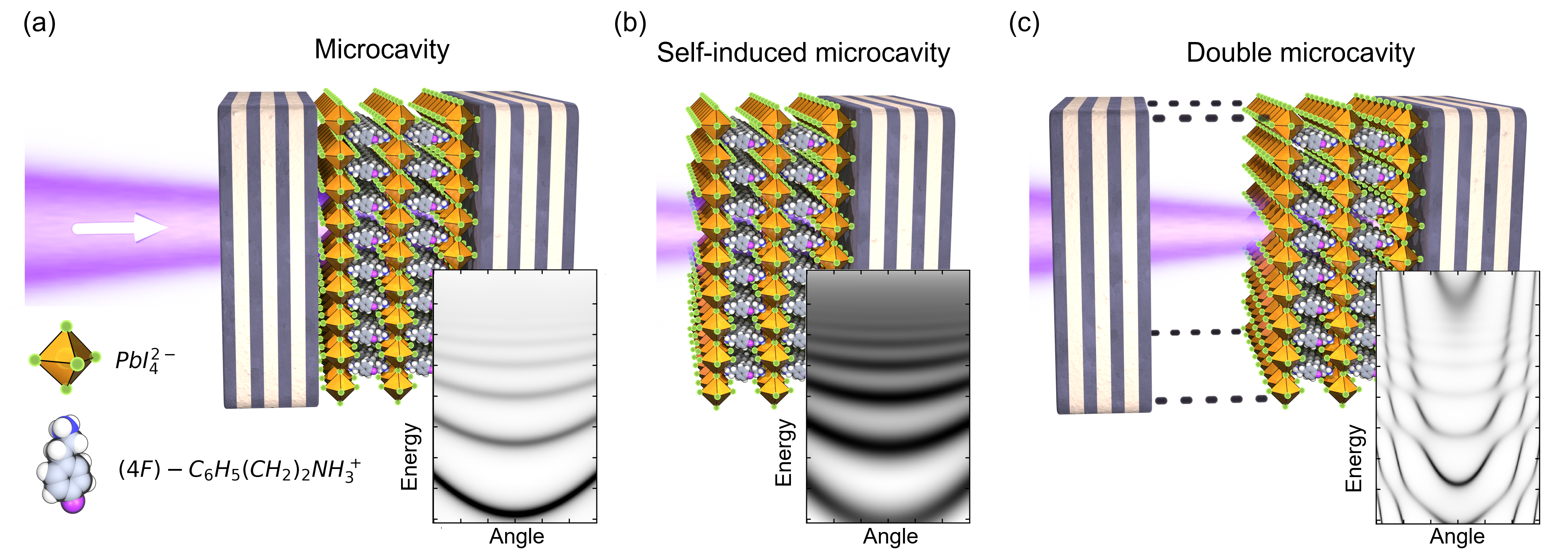}
\caption{Schematic explanation of the differences between (a) microcavity (b) self-induced microcavity and (c) double microcavity. The insets show schematic simulations of the respective dispersion relations of photonic modes in these situations. 
\label{fig1}}
\end{figure*}

\section{Introduction} 
Specialized design of microscale optical structures for dispersion engineering in the strong light-matter regime has recently provided excellent means of exploring polaritonic band topology~\cite{Gianfrate_Nature2020}, artificial gauge fields~\cite{Lim_NatComm2017}, and non-hermitian physics~\cite{Su_SciAdv2021}. In contrast to two-dimensional photonic crystal slabs~\cite{Zhen_Nature2015, Zanotti_PRB2022}, engineering the photon dispersion in one-dimensional photonic crystals, or planar microcavities, is more limited unless exotic materials are introduced which can help mix the polarization states of the confined photon~\cite{Rechcinska_Science2019, Kokhanchik_PRB2021}, mix light and matter~\cite{Basov_Nanopho2021, Luo_ApplPhysRev2023}, or both~\cite{Lundt_NatNanotech2019, Lempicka_SciAdv2022}. Through the loss of strong coupling, it is also possible to observe exceptional points \cite{Opala:23}. 

Ruddlesden–Popper perovskites (RPPVs) represent a unique class of materials for the applications in polaritonics~\cite{Lanty_2008, Bloch_2012, Su_NatMat2021}. The high binding energy of multilayer excitons therein allows the observation of exciton-polaritons (hereafter referred to as polaritons) up to room temperature. In our investigation, we used 4-fluorophenethylammonium lead iodide, (4F-PEA)$_2$PbI$_4$ (hereafter denoted as PEPI-F), known for its good photostability\cite{Coriolano}. We employed an open microcavity~\cite{Schneider2023} which offers an accessible and highly adjustable platform to observe the effects of strong photon-exciton coupling~\cite{Dufferwiel_APL2014, Krol_OMExp2023}. The ability to control the distance between DBRs in open cavities also allows us to form an air gap that we used for non-Hermitian polariton-photon coupling in this paper. 

Typically, to observe strong light-matter coupling the perovskite material is sandwiched between DBRs, as schematically illustrated in Figure \ref{fig1}(a), forming a so-called extrinsic cavity. The inset shows the simulated reflectivity spectra by a transfer matrix method (the details of the method are provided further on in the text). Strong coupling can also be obtained without extrinsic confinement of light, with the perovskite crystal itself acting as an intrinsic Fabry-Pérot resonator if the crystalline facets are of high quality~\cite{Han_ACSPho2020, Anantharaman_NanoLett2021}. Figure \ref{fig1}(b) illustrates such a case with perovskite crystal deposited on one DBR mirror and the cavity is formed by crystal surface.  

However, the polariton dispersion gets drastically modified if instead two such cavity concepts are combined together with an additional air gap, to form a hybrid extrinsic-intrinsic cavity system, as illustrated in Figure \ref{fig1}(c) [i.e., an off-centered cavity system]. In such an inversion-symmetry broken double-cavity system the interaction between the modes of these two cavities results in polaritons characterized by a unique dispersion with multiple inflection points, as illustrated in the inset, and observed in the absorption and emission spectra. The unusual multi-modal dispersion relation in conjunction with the exciton-photon detuning dependent polariton decay rates through the broad exciton absorption peak implies a complicated multiparameter non-Hermitian mechanism. It leads to the natural loss of strong coupling in polariton modes with high exciton fraction, thereby providing an opportunity to investigate the physics of exceptional points~\cite{Miri_Science2019, Li_NatNanotech2023}. As a result of the large value of the Rabi splitting the 2D RPPVs such as PEPI-F, the loss of strong coupling is not effective. Nevertheless, by coupling polariton modes with air-gap photons, we are able to observe this effect due to much lower intermode coupling. 
 
This effect, together with the possibility of replacing air with various materials in the open microcavity system, expands the range of possible applications of these structures. Specifically, such novel exceptional points created can be applied in the fields of sensor technology~\cite{Chen2017, Wiersig2014, Hodaei2017} and for control over loss and gain for lasing phenomena~\cite{Peng2014, Gao2022}. 

\begin{figure*}
\centering
\includegraphics[width=\textwidth]{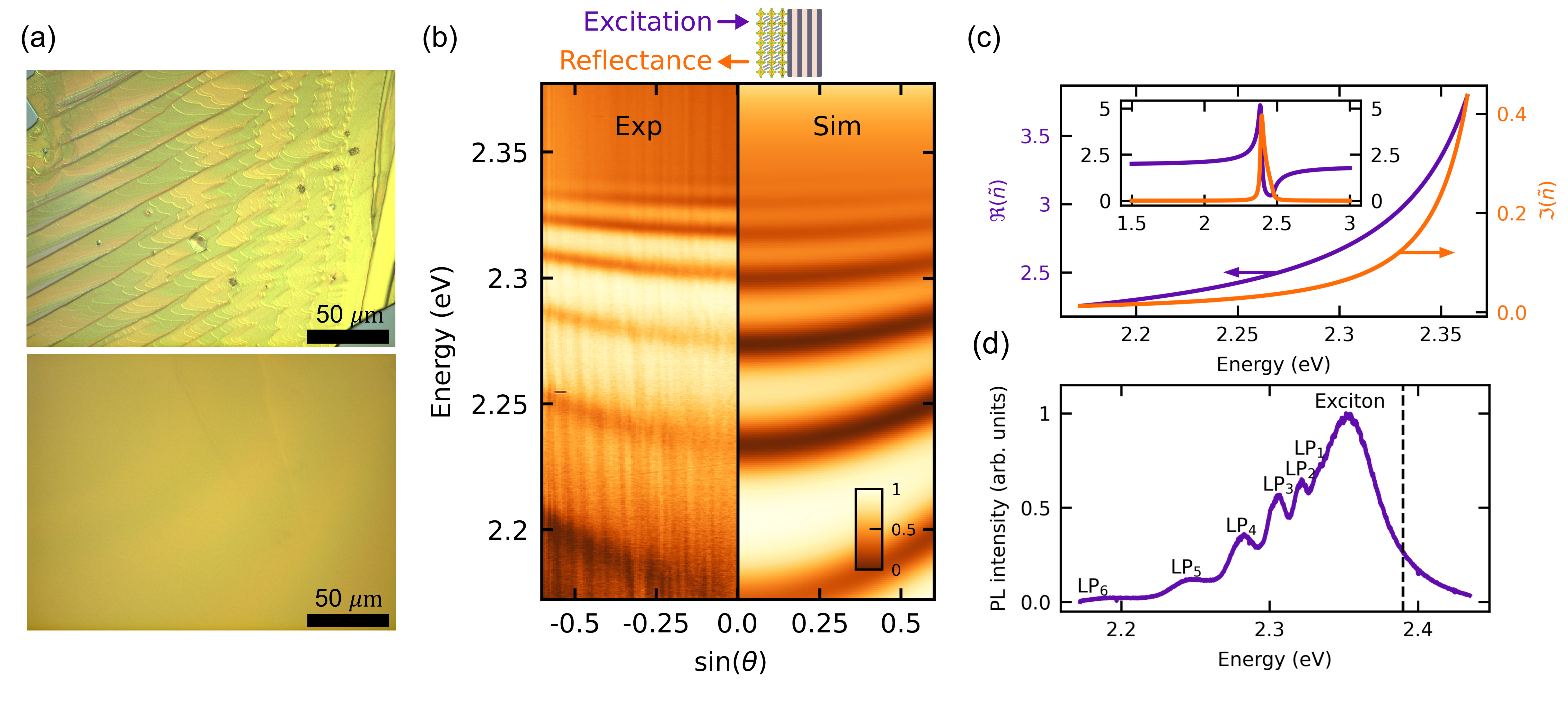}
\caption{Panel (a) shows a Nomarski contrast microscope image of PEPI-F monocrystals in two situations: multilayer structure (top) and high quality homogeneous thickness crystal used in our study (bottom), (b) experimental results (left side) and simulation (right side) of normalized white light reflection for self-induced microcavity with PEPI-F, as schematically shown in the inset above the spectra and in Figure \ref{fig1} (b). Panel (c) shows fitted real and imaginary parts of the complex refractive index. Inset: in the wider spectral range. The photoluminescence spectrum at normal incidence of the self-induced microcavity is shown in panel (d). Polariton peaks and broad exciton emission are marked. The dashed line marks the position of the exciton in PEPI-F.
\label{fig2}}
\end{figure*}

\section{Results}
\subsection{Self-induced perovskite microcavities}
For the observation of polaritons in the intrinsic self-induced perovskite microcavity, it is essential to fabricate high-quality perovskite monocrystals, rather than polycrystalline or nanoparticle layers, to obtain perfect crystal facets. The synthesis of layered PEPI-F monocrystals using the Anti-solvent Vapor-assisted Capping Crystallization method \cite{Ledee2017} was conducted directly on the substrate, composed of a DBR with 6 TiO$_2$/SiO$_2$ pairs.  For a more detailed description of the synthesis, see the Supporting Information (SI). The layered structure of quasi-2D RPPVs, shown in the top part of Figure \ref{fig2}(a), allows the adjustment of the crystal thickness through mechanical exfoliation, eliminating the need for spatial confinement crystallization. In our case, the crystal was exfoliated to a thickness of 1.4~$\mu$m. The uniformity of the surface, after the exfoliation process, can be seen in the lower part of Figure \ref{fig2}(a).

To investigate the optical properties of the PEPI-F perovskite crystal deposited on DBR, we conducted both reflectance and nonresonant photoluminescence (PL) measurements in reflectance configuration. Both experiments were performed at room temperature, the signal was detected in the horizontal polarization of emitted/reflected light (for detection in vertical polarization, a similar effect can be observed, albeit shifted in energy due to the birefringence of PEPI-F). 

 The presence of several distinct parabolic shaped bands in the angle-resolved white light reflectance spectra, depicted in the left half of Figure \ref{fig2}(b), which become less-dispersive closer to the exciton line at $E_X = 2.39$ eV confirms efficient photon confinement and the emergence of polaritons within the perovskite crystal. This confinement and strong light-matter coupling is facilitated by the high refractive index of the perovskite crystal, leading to internal reflection of light at the interface between the crystal and the environment. 

We assume that, within the investigated spectral range, the dielectric function of the perovskite can be sufficiently well described by the Lorentz model:
\begin{equation}
\label{eq:1}
\epsilon(E) = \epsilon_{\infty}+\frac{f}{E_X^2-E^2-i\gamma E}.
\end{equation}
Here, the dielectric constant at infinity $\epsilon_{\infty} = 3.57$, oscillator strength $f = 1.51$ (eV)$^2$, and damping coefficient $\gamma=  0.017$ eV were determined from reflection measurements. Exciton resonance energy $E_X = 2.386$ eV was taken from previous study performed on the same material \cite{Polimeno2021}. The obtained calculated reflectivity map using thetranfer matrix method~\cite{Berreman_Optica72} is depicted in the right half of Figure \ref{fig2}(b). The corresponding real and imaginary parts of the complex refractive index ($\tilde{n} = \sqrt{\epsilon}$) are presented in Figure \ref{fig2}(c). 

\begin{figure*}
\includegraphics[width=\textwidth]{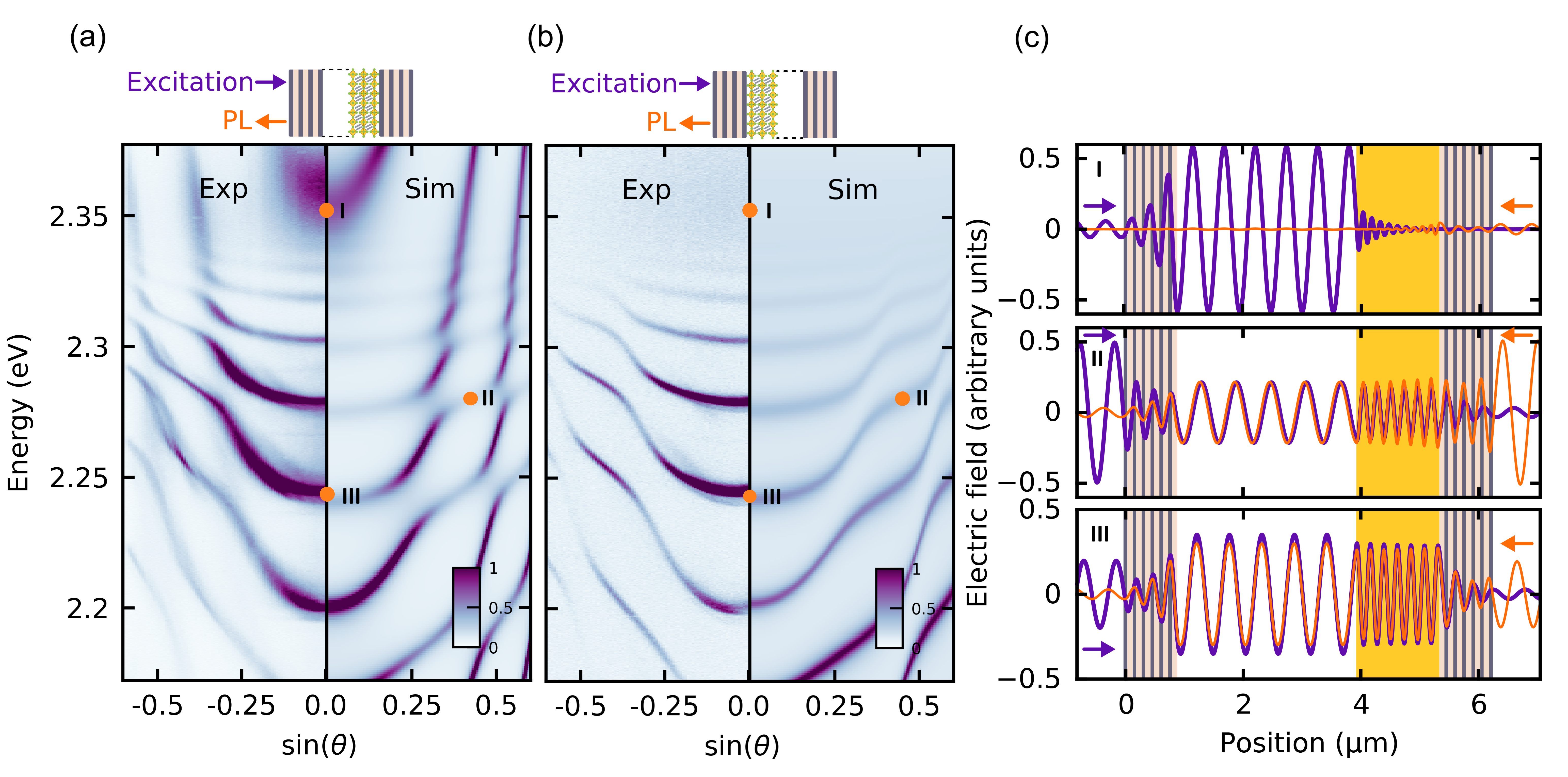}
\caption{Panels (a,b) show the experimental results (left sides) and simulations (right sides) of perovskite double microcavity emission. Panel (a) is for excitation from the air gap side, and panel (b) is for excitation from the perovskite side. Panel (c) show the electric field distributions inside the cavity for forward (violet) and backward (orange) excitation, respectively. The energies and sines of the collection angle for the panels from top to bottom are (2.355 eV,0), (2.275 eV,0.4) and (2.245 eV,0), respectively and are marked with orange dots and numbers on panels (a,b).
\label{fig3}}
\end{figure*}

The sample was excited using non-resonant continuous-wave laser with a wavelength of 405~nm. The measured PL emission spectra at normal incidence is shown in Figure \ref{fig2}(d), revealing a pronounced broadband peak around the excitonic resonance energy, a characteristic effect of hybrid organic-inorganic metal halide perovskites at room temperature. Additional lower-polariton peaks below the exciton line can also be seen.

\subsection{Double microcavity}
After characterizing the polariton features of the intrinsic perovskite-DBR cavity, we next move onto the hybrid extrinsic-intrinsic cavity setup shown schematically in Figure \ref{fig1}(c). Here, an open microcavity was constructed with a second identical DBR, and the thickness of the air gap between the perovskite crystal and the top DBR was controlled using a piezoelectric element. This setup forms an unconventional double microcavity, as confined electromagnetic modes in the air gap and the perovskite layer couple through the air-perovskite interface. 
The presence of the additional `empty' cavity (i.e., air gap between DBR and perovskite) can be described as coupling between perovskite polariton modes with a new family of photon modes in the air gap. In the SI we provide simulations illustrating how the addition of an air cavity influences the polariton dispersion.

Figure \ref{fig3}(a) displays the results of PL measurements in reflection configuration for a microcavity composed of the crystal from Figure \ref{fig2} and an air gap of approximately 3~$\mu$m. The air gap was located on the excitation side, while the perovskite crystal was positioned at the back, as depicted in the schematic inset in Figure \ref{fig3}(a). The resultant polaritons exhibit an unusual non-quadratic dispersion relation with a rapidly changing effective mass (concavity) separated by more-than-one inflection point. This is in sharp contrast to standard planar cavity polaritons which usually display only a single lower polariton branch with a single inflection point~\cite{Su_NatMat2021}.

However, as the exciton absorption of the perovskite is spectrally broad, and significantly increases at higher energies (with the maximum at the exciton resonance), a reduction of the coupling strength between polariton modes and external (i.e. from air cavity) photon becomes apparent, visible as a decrease in the anti-crossing of subsequent polariton modes. This ultimately leads to a loss of strong coupling, as we explain in more detail in the Model section. Our measurements are well reproduced with the numerical transfer matrix method shown to the right in Figure \ref{fig3}(a). The difference in the intensity (i.e. population) of the modes observed between simulation and experiment is due to the fact that the distribution of PL intensity is affected by carrier relaxation processes in the perovskite which is not captured in the Berreman method.

\subsection{Asymmetric spectrum}
Because of the broken inversion symmetry (or reflection symmetry along the optical axis) the observed emission spectrum depends on which side of the double cavity is excited, as illustrated in the top insets of Figures~\ref{fig3}(a) and~\ref{fig3}(b). Notably, the spectrum is clearly affected by the strong absorption of the incident light at the perovskite layer. This asymmetry is also well visible in the calculated electric field distribution along the cavity, as illustrated in Figure \ref{fig3}(c) for the excitation from the side of the air-gap (purple solid line) and the perovskite (orange solid line). These observations confirm a non-reciprocal response of the double-cavity towards left and right incident light, exhibiting different mode occupations in the polariton branches.

For example, looking at the orange dot marked ``I'' around 2.35 eV in Figure \ref{fig3}(a), we predict a polariton made up of high external photon fraction in the emission spectrum as seen from the dominant purple electromagnetic field profile in the top panel of Figure~\ref{fig3}(c). Conversely, when our system was excited from the perovskite side, Figure~\ref{fig3}(b), the emission at ``I'' vanishes because of the smaller photonic component inside the structure. The external photonic mode is negligible in the spectrum due to the pronounced light absorption in the perovskite, as shown in Figure \ref{fig3}(c) in the first panel by the orange curve.
The electric field distribution inside the cavity for two other energies and emission angles marked ``II'' and ``III'' are illustrated in Figure~\ref{fig3}(c) in the middle and bottom panels. For far detuned modes the asymmetric field distribution is less pronounced, but the significant difference in the mode amplitude is visible for both normal incidence (bottom panel) and high angle (middle panel). 

We observe that resonant photons are not anticipated in an air gap cavity due to the substantial energy difference between laser energy and the external photonic mode. 
However, we observe experimentally a nonzero occupation of external photonic modes. This phenomenon is plausible because resonant photons can be introduced into the air gap cavity through perovskite exciton emission.
Depending on the excitation direction, photons from the air cavity can be either effectively or ineffectively populated. 

\section{Model} 
Based on experimental observations, we have developed a phenomenological theory describing the properties of perovskite polaritons coupled to the air gap photon field in the double perovskite microcavity.
 
The starting point of our investigation involves only the multilayer perovskite crystal in which excitons and intrinsically trapped photons strongly coupled. In their recent work, Mandal et al.~\cite{Mandal_2023} have demonstrated that polaritons in filled cavities can be described by a $2N \times 2N$ Tavis-Cummings Hamiltonian, where each of the $N$ photonic modes is coupled to a distinct exciton ensemble. Extending the proposed approach to our hybrid air-perovskite double cavity we include important non-Hermitian properties of the system by accounting for the different lifetimes of photons and excitons within the perovskite crystal. We then introduce coherent coupling between the perovskite crystal polariton modes and the external air gap photon field. For this we propose an effective Hamiltonian, ensuring agreement with both experimental observations and the results obtained from Berreman simulations. Our results suggest that exceptional points can be realised in our system by appropriately tuning external air-gap photon energies into resonance with the perovskite polariton branches.

\subsection{ Perovskite crystal cavity} 

First, we consider the effective Hamiltonian matrix describing polariton modes in a perovskite crystal placed on the distributed Bragg reflector, schematically shown in Figure~\ref{fig1}(b). In our approach, we assume that the following complex energies can describe the confined photonic modes in the perovskite crystal:
\begin{equation}
\label{eq:2}
\tilde{E}^{(n)}_{C}({\bf{k}}) = E_{C,0}^{(n)} + \frac{\hbar^2 {\bf{k}}^2}{2 m^*_C} - i\hbar\gamma_C,
\end{equation}
where $E_{C,0}^{(n)}$ corresponds to the energy of the $n$-th mode at $k=0$, and $\mathbf{k}\equiv (k_x,k_y)^\text{T}$ is the in-plane wavevector. The wavevector is related to the emission angle presented in Figures \ref{fig2} and \ref{fig3} through $k = \tilde{E}_C/\hbar c \sin \theta$ where $c$ is the speed of light and $\hbar$ is the Planck constant. The effective mass of the photon inside the crystal is given by
$m_C^*$ (parabolic approximation), and their decay rate, inversely proportional to their lifetime, by $\gamma_C$. In the above equation and the rest of the work, we use a tilde superscription to highlight the nonzero complex part of any physical observable. Since the substantial difference in effective masses between excitons in the perovskite crystal, denoted as $m^*_X$, and cavity photons, the energies of excitons can be treated as wavevector independent:
\begin{equation}
\label{eq:3}
 \tilde{E}_X=E_{X,0}-i\hbar\gamma_X.
\end{equation}
 where $E_{X,0}$ denote exciton resonance and parameter $\gamma_X$ is the exciton decay rate. 
 
 Strong coupling between excitons and photons, denoted by the Rabi frequency $\Omega_R$, gives rise to new polariton eigenmodes in the crystal. The investigated energy range of perovskite modes is smaller than the characteristic Rabi energy, which allows us to assume same $\Omega_R$ for all modes (i.e., the relevant modes are close to the exciton line). Moreover, the layered structure of the perovskite material, which includes multiple quantum wells, facilitates effective coupling between the photonic modes.

 Consequently, the polariton modes arising in the perovskite crystal (PC) can be described by the following block Hamiltonian: 
\begin{equation}
H_\text{PC}=\begin{pmatrix}
\tilde{E}_C^{(1)}({\bf{k}}) & \frac{1}{2}\hbar\Omega_R & 0 & 0 & 0 \\
\frac{1}{2}\hbar\Omega_R & \tilde{E}_X({\bf{k}}) & \cdots  & 0 & 0  \\
0 & \vdots &  \ddots & \vdots & 0  \\
0 & 0 & \cdots & \tilde{E}_C^{(n)}({\bf{k}}) & \frac{1}{2}\hbar\Omega_R  \\
0 & 0& 0 & \frac{1}{2}\hbar\Omega_R & \tilde{E}_X({\bf{k}}) \\
\end{pmatrix}.
\label{eq:4}
\end{equation}
Through the diagonalization of the matrix described above, we obtain $n$ distinct lower ($LP$) and upper ($UP$) polariton dispersion branches, each possessing the following energy eigenvalue:
\begin{equation}
\label{eq:5}
    \tilde{E}_{UP,LP}^{(n)}({\bf{k}})=
    \bar{E}_n\pm\frac{1}{2}\sqrt{\delta_{n}^2+\hbar^2\Omega_R^2}-i\Gamma_{UP,LP}^{(n)},
\end{equation}
where $\bar{E}_n$ and $\delta_n$ define respectively the average and difference between the real part of the $n$-th photon mode and exciton energies (wavevector-dependent). The last parameter of the right-hand side of the above equation, $\Gamma_{UP,LP}^{(n)}$, defines the effective decay rate of upper and lower polariton branches (distinct by $UP$ and $LP$, respectively). For lower polariton modes, the imaginary part of dispersion is given by:
 
 \begin{equation}
 \label{eq:6}
\Gamma_{LP}^{(n)}({\bf{k}})=\gamma_X|X_n({\bf{k}})|^2+\gamma_C|C_n({\bf{k}})|^2,
\end{equation}
 where $|X_n|^2$ and $|C_n|^2$ are the exciton and photon Hopfield fractions of the $n$-th dispersion branch, which fulfill the normalization condition $|X_n({\bf{k}})|^2+|C_n({\bf{k}})|^2=1$, and define the polariton composition.  In the limit of small losses, their values are directly related to the energy detuning between the real part of photon and exciton energies~\cite{Hopfield_1958,Deng_2010}:
\begin{equation}
\label{eq:7}
X_n({\bf{k}})=\frac{1}{\sqrt{2}}\bigg(1+\frac{\delta_n({\bf{k}})}{\sqrt{\delta_n({\bf{k}})^2+\hbar^2\Omega_R^2}}\bigg)^{\frac{1}{2}},
\end{equation}
\begin{equation}
\label{eq:8}
C_n({\bf{k}})=\frac{1}{\sqrt{2}}\bigg(1-\frac{\delta_n({\bf{k}})}{\sqrt{\delta_n({\bf{k}})^2+\hbar^2\Omega_R^2}}\bigg)^{\frac{1}{2}}. 
\end{equation}
\begin{figure}[t]
\centering
\includegraphics[width=0.8\textwidth]{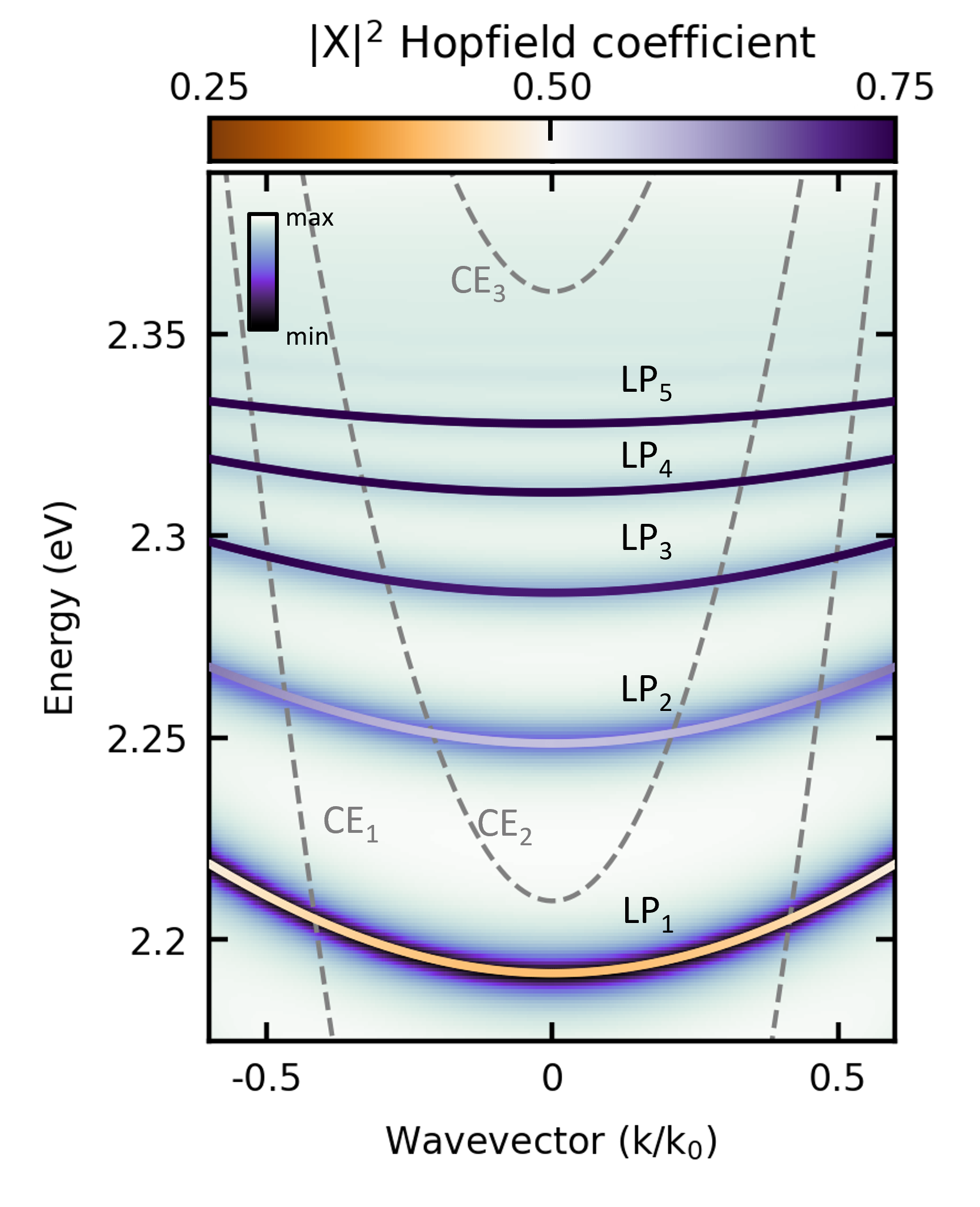}
\caption{  The dispersion of lower polaritons in a perovskite crystal placed on the DBR is calculated using diagonalization techniques. Bold lines represent the real part of the lower polariton branches, determined using the dissipative version of the Tavis-Cummings Hamiltonian. The colour of the lines is determined by the excitonic Hopfield coefficient $|X|^2$, reflecting the excitonic fraction of the calculated modes. The background colourmap corresponds to the simulation conducted using the Berreman method. The dashed lines show the location of the pure photonic modes $CE_1$, $CE_2$, and $CE_3$ in the external air gap cavity. The above emission and dispersion are plotted using a normalized wave vector, where $k_0$ is the scaling parameter.}
\label{fig:3new}
\end{figure}
Due to the fact that in perovskite crystals $\hbar\Omega_R \gg \delta(\mathbf{k})$ we can neglect the weakly illuminated upper polariton branches above the exciton line. Therefore, we truncate our state space around the lower polariton energies:
\begin{equation}
\label{eq:9}
H_{LP}^\text{PC}=\begin{pmatrix}
\tilde{E}_{LP}^{(1)}({\bf{k}}) & 0 & 0  & 0 \\
0 &\tilde{E}_{LP}^{(2)}({\bf{k}}) & \cdots   & 0  \\
0 & \vdots &  \ddots & \vdots  \\
0 & 0& \cdots  & \tilde{E}_{LP}^{(n)}({\bf{k}}) \\
\end{pmatrix}.    
\end{equation}
We fit the parameters of the above Hamiltonian to obtain the best correspondence between their eigenenergies and the polariton dispersion calculated using the Berreman method. Parameters and detailed descriptions of the fitting procedure are included in the SI. Figure \ref{fig:3new} shows the fitted dispersion relations for five perovskite polariton modes. Each of the presented modes is characterized by a different light-matter composition, affecting their effective mass and decay rate. The colour scale in Figure \ref{fig:3new} shows the exciton fraction of each branch, where light orange and deep purple correspond more photonic and excitonic modes, respectively. The Hopfield coefficients indicate that the lowest energy LP branch is mainly photonic, while the other modes have a dominant exciton contribution. Notably, the decay rate for excitons in perovskite crystal at room temperature $\gamma_X$ is substantially greater than that for photons $\gamma_C$, which leads to faster dissipation for polariton modes closer to the exciton line. 

\begin{figure}[t]
\centering
\includegraphics[width=0.8\textwidth]{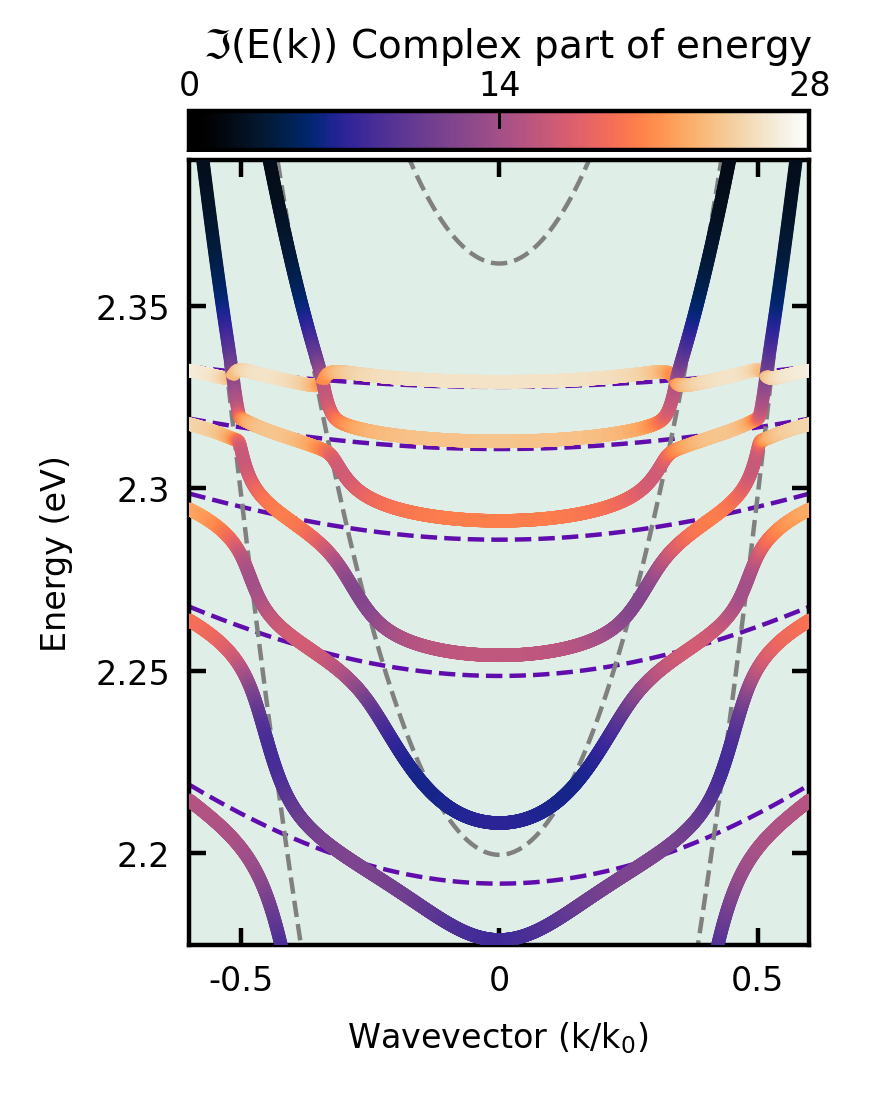}
\caption{ The dispersion of lower polaritons coupled with the external photon modes calculated using coupled mode theory. The bold line illustrates the dispersion of double cavity modes, with the colour representing the complex part of energy (given in meV) associated to the dissipation. The grey and purple dashed lines describe the dispersion of external photons and polaritons, respectively.}
\label{fig:4}
\end{figure}

\subsection{Double self-inducted perovskite cavity} 
Hamiltonian~(\ref{eq:4}) characterizes polariton modes only within the perovskite crystal. For the double-cavity configuration depicted in Figure~\ref{fig3}(c), 
we treat the presence of $N_E$ external air-gap photon modes perturbatively through coupled mode theory to account for their influence in the perovskite layer. The complex energies associated with these air-gap modes can be represented as:
\begin{equation}
\label{eq:10}
\tilde{E}^{(l)}_{E}({\bf{k}}) = E_{E,0}^{(l)} + \frac{\hbar^2 {\bf{k}}^2}{2 m^*_{E}} - i\hbar\gamma_{E}.
\end{equation}
Based on experimental observations and coupled mode theory, we propose a Hamiltonian describing the interaction between lower polariton perovskite modes and the external cavity fields:
\begin{equation}
\label{eq:11}
H=\begin{pmatrix}
\tilde{E}_{E}^{(1)}({\bf{k}})  &0& J_C^{(1,1)} &  \cdots  & J_X^{(1,n)} \\
0 &\tilde{E}_{E}^{(2)}(\bf{k}) &J_C^{(1,2)} & \cdots   & J_X^{(2,n)} \\
J_C^{(1,1)} &J_C^{(1,2)} &\tilde{E}_{C}^{(1)}({\bf{k}}) &\cdots & 0    \\
{J_X^{(1,1)}} &J_X^{(1,2)} &\frac{1}{2}\hbar\Omega_R &\cdots & 0    \\
 \vdots & \vdots & \vdots &  \ddots & \vdots  \\
J_X^{(1,n)} &J_X^{(2,n)} & 0& \cdots  & \tilde{E}_{X}^{(n)}({\bf{k}}) \\
\end{pmatrix},    
\end{equation}
where $J_X^{(n,l)}$ and $J_C^{(n,l)}$ are parameters describing the interaction between external air-gap photonic modes and excitons and perovskite crystal cavity modes, respectively. In the considered case, only two external photon modes coupled to the polariton branches are relevant and therefore we limit ourselves to $l=2$ (see $CE_1$ and $CE_2$ in Figure \ref{fig:3new}. We tuned the parameters of our model, such as external photonic fields and the strength of phenomenological interaction constants, to reproduce the experimentally observed dispersion curves. Results are shown in Figure~\ref{fig:3new}. It should be noted that due to the increasing excitonic fraction of polaritons belonging to the lower polariton modes of higher energy (LP3, LP4, LP5), we can observe the closing of the band gap between lower polariton branches and external cavity modes. This effect is connected to the non-Hermitian nature of the considered system, in which exciton losses dominate the coupling strength (higher complex energy part). In such a system, exceptional points can arise.

\subsection{Exceptional points in a double cavity}
In recent years, exceptional points in the exciton-polariton system have been extensively studied from experimental and theoretical perspectives. The unique properties of exciton polaritons make it possible to observe such non-Hermitian features in experiments employing polarization degree of freedom \cite{Gao2018,Król2022,Su_SciAdv2021,Liao2021,Li2022}, double-well traps [6], perovskite-based non-local or plasmonic metasurfaces \cite{Masharin_2023, Lu_2020} or chaotic non-Hermitian billiards \cite{Gao_2015} Recent works \cite{Khurgin_2020, Opala:23, Rahmani_2024} highlight that exceptional points are a natural feature of coupled light-matter systems at the crossover between coupling regimes. This work extends the ideas of the mentioned works to an open perovskite cavity, where gain and dissipation can be engineered through the mode resonance position in the energy spectrum. Compared to previous works, the considered system offers advantages in generating multiple resonances (or inflection points) within the polariton-photon spectrum. Even within the same dispersion branch, these resonances can exhibit fundamentally different physical properties caused by trivial mode crossings and exceptional points.

To explore the possibility of observing exceptional point (EP) in our double microcavity setup, we consider a simplified model considering the interaction between the external air-gap photon mode and a single perovskite polariton branch,
\begin{equation}
\label{eq:12}
H_{SB}=\begin{pmatrix}
 {E}_E({\bf{k}})-i\hbar{\gamma}_E & J_\text{eff}({\bf{k}})  \\
J_\text{eff}({\bf{k}}) & E_{LP}({\bf{k}})-i\hbar\Gamma_{LP}({\bf{k}})
\end{pmatrix}.
\end{equation}
From here on, we will drop the subscript $LP$ for simplicity and define $\Gamma_{LP} \equiv \Gamma$. Here $J_\text{eff}({\bf{k}})$ defines the coupling between lower polariton and external cavity modes. This approach becomes accurate if different perovskite polariton modes are well separated in energy compared to the strength of the coupling $J_\text{eff}({\bf{k}})$. The mentioned coupling is dependent on the excitonic and photonic fractions constituting a lower polariton state:
\begin{equation}
    J_\text{eff}({\bf{k}})=J_C|C({\bf{k}})|^2+J_X|X({\bf{k}})|^2.
\end{equation}
The eigenmodes of the above Hamiltonian have complex energies:
\begin{equation}
\begin{split}
E^{\pm}_{CP}({\bf{k}})= \frac{1}{2}({E}_E({\bf{k}})+E_{LP}({\bf{k}})-i\hbar({\gamma}_E+\Gamma({\bf{k}})) \\
\pm\sqrt{[{E}_E({\bf{k}})-E_{LP}({\bf{k}})-i\hbar({\gamma}_E-\Gamma({\bf{k}}))]^2+4J_\text{eff}({\bf{k}})^2}. 
\end{split}
\end{equation}
Analysing the equation above, two fundamentally different scenarios can be distinguished. First, when $J_\text{eff}({\bf{k}})>\frac{\hbar}{2}|{\gamma}_E-\Gamma({\bf{k}})|$, polaritons are strongly enough influenced by the additional external photon field to reveal characteristic anticrossing at the resonance between $\tilde{E}_C({\bf{k}})$ and $E_{LP}({\bf{k}})$. Second, when $J_\text{eff}({\bf{k}})<\frac{\hbar}{2}|{\gamma}_E-\Gamma({\bf{k}})|$, 
the lower polariton and external photon modes are weakly coupled with only a
small admixture of the other. Both of the mentioned scenarios are distinguished by exceptional points observed when:
\begin{equation}
\label{eq:15}
J_\text{eff}({\bf{k}})=\frac{\hbar}{2}|{\gamma}_E-\Gamma({\bf{k}})|.
\end{equation}
Quite recently, exceptional points were tracked to the difference between a many body phase transition into a polariton condensate or photon lasing regime~\cite{Hanai_PRL2019}. It should be noted that the parameters $\Gamma$ and $J_\text{eff}$ are wavevector-dependent and can vary across the polariton dispersion. Therefore, the appearance of the exceptional point strongly depends on where the modes cross (or anticross) in the dispersion because of their varying decay and coupling strength.

\begin{figure}[t]
\centering
\includegraphics[width=0.8\textwidth]{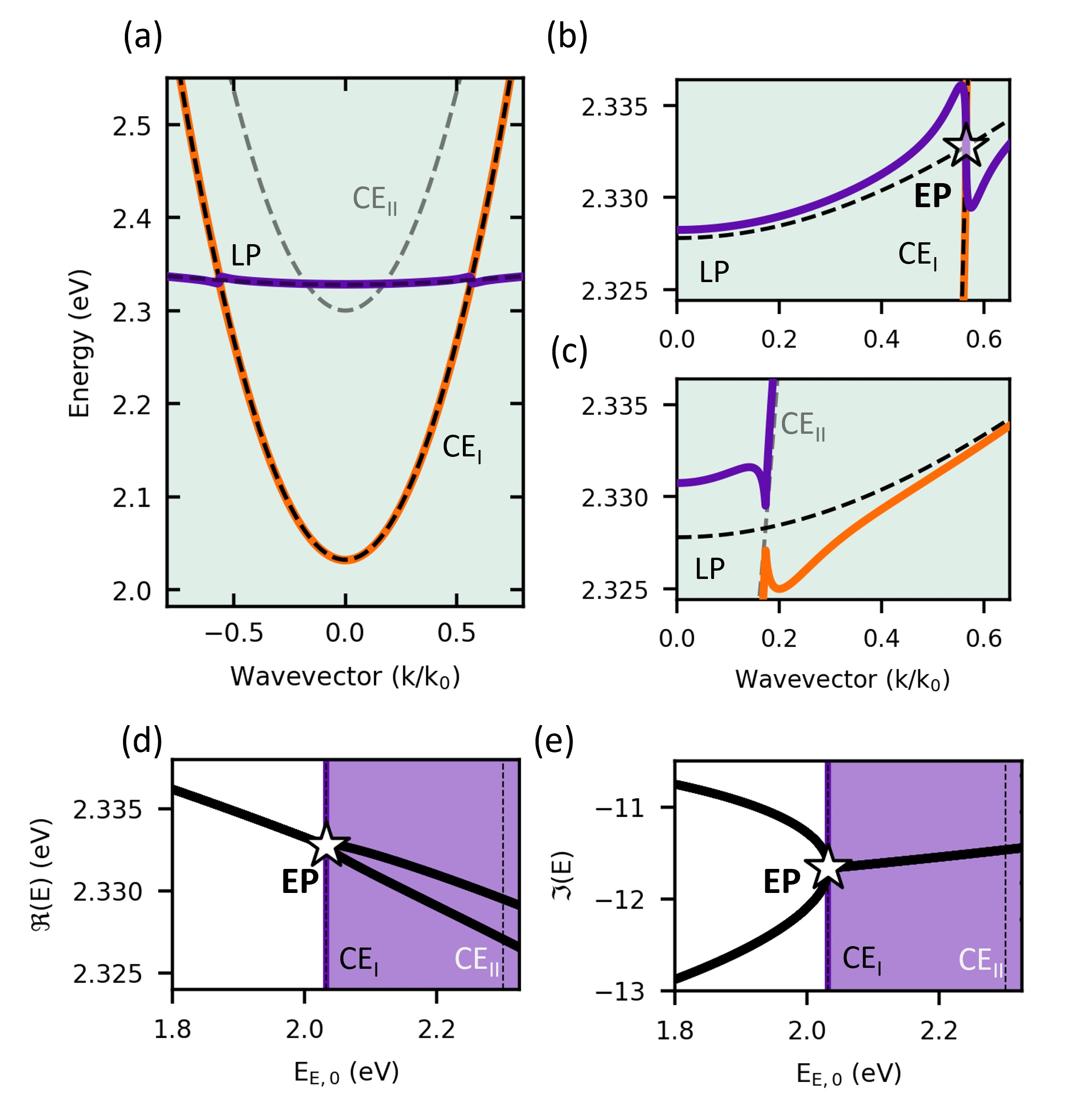}
\caption{Exceptional Point in Double Self-Induced Perovskite Microcavity. Panel (a) displays the real part of the eigenvalues of Hamiltonian~(\ref{eq:12}), represented by deep purple and orange lines. The black and grey dashed lines correspond to two external photon modes. Panel (b) illustrates the resonance between the lower polariton and mode $CE_I$. This specific point in the energy spectrum satisfies all the conditions for the emergence of an exceptional point. In contrast, the opposite situation is presented in panel (c). In this case, the system remains in a strong coupling regime and exhibits a characteristic anticrossing between $CE_{II}$ and the polariton modes. Panels (d) and (e) present phase diagrams that represent the real and imaginary parts of the energy of the considered Hamiltonian as a function of the minimum of the external photon mode, described as $E_{E,0}$, at the resonance. The star highlights the point in the parameter space where exceptional points emerge.}
\label{fig:6}
\end{figure} 

To demonstrate the appearance of exceptional points in the double-cavity system, we calculate the eigenspectrum of the reduced Hamiltonian~(\ref{eq:12}) and present the results in Figure \ref{fig:6}.  Here, we consider a potential scenario in which the significant decay of excitons leads to the closure of the bandgap between external photons and polariton modes. The dispersion shown in Figure~\ref{fig:6}(a) displays the eigenvalues of Hamiltonian~(\ref{eq:12}) containing exceptional points (\ref{eq:15}) around the resonance between the two branches. Panel (b) provides a zoomed-in view of the considered dispersion near the (right-hand side) exceptional point, depicted as a transparent star.

It should be noted that the closure and opening of the bandgap, as well as the passage through the exceptional point, can be achieved by tuning the resonance wavevector. Such a situation is illustrated in panel (c), where the coupling between the lower polariton and the mode, denoted as $CE_{II}$ modes, can occur because, for this resonance, the coupling is higher than the losses in the system. We performed a numerical simulation to demonstrate the occurrence of exceptional points concerning the detuning of the external photon mode from the exciton energy. Panels (d) and (e) in Figure 6 illustrate both the real and imaginary components of the eigenenergies obtained through the diagonalization of the Hamiltonian. The parameter $E_E$ represents the energy of the external air-gap photon mode.

The experimental observation of the exceptional point requires a unique sample design, allowing a smooth transition between regimes separated by condition (15). This can be realized by wedged cavity or precise tuning of the cavity thickness. Nevertheless, based on our theoretical study, such points can arise in the analyzed light-matter spectrum. This observation can serve as motivation for further investigation.

\section{Conclusions} 

In summary, we constructed a double microcavity system, integrating two-dimensional Ruddlesden–Popper perovskites, which enable the exploration of non-Hermitian phenomena. The combination of an air microcavity with a self-formed perovskite microcavity results in a unique dispersion exhibiting asymmetric features. These characteristics can be consistently replicated using both numerical transfer matrix methods and a non-Hermitian coupled oscillator model. Additionally, incorporating the perovskite's inherent absorption losses leads to observable phenomena such as the closing of bands, a consequence of the loss of strong coupling. Theoretically, our analysis shows emergence of an exceptional point within the system. This offers new perspectives on the interplay between material properties and photonic structures in the study of non-Hermitian physics.

\subsection*{Funding}
This work was supported by the National Science Center, Poland, under the projects: 2022/47/B/ST3/02411, 2021/43/B/ST3/00752, 2019/35/N/ST3/01379. This work was financed by the European Union EIC-Pathfinder project 'Quantum Optical Networks based on Exciton-polaritons" (Q-ONE, Id: 101115575). H.S. acknowledges the project No. 2022/45/P/ST3/00467 co-funded by the Polish National Science Centre and the European Union Framework Programme for Research and Innovation Horizon 2020 under the Marie Skłodowska-Curie grant agreement No. 945339. A.O. acknowledges support from the Foundation for Polish Science (FNP).

\subsection*{Author contributions}
M.K., M.Kr, B.P. and A.O. conceived the idea, A.O., and P.K. developed the theoretical description, M.K., and M.Kr. performed the optical experiments and numerical simulations, R.M. and W.P. grew dielectric mirrors, M. K. grew perovskite crystals, M.K., A.O., H.S., and B.P. wrote the manuscript with input from all other authors, J.Sz., M.M., H.S., A.O. and B.P. supervised the project.

\subsection*{Conflict of interest}
Authors state no conflict of interest.

\subsection*{Data availability statement}
The datasets generated during and/or analyzed during the current study are available from the corresponding author on reasonable request.


\renewcommand{\theequation}{S\arabic{equation}}
\renewcommand{\thefigure}{S\arabic{figure}}
\renewcommand{\thesection}{S\arabic{section}}

\section*{Supporting Information}

\subsection*{Crystalisation of PEPI-F}
Synthesis of the solution for PEPI-F crystallization was carried out in an argon-filled glovebox by mixing PbI$_2$ and (4F)-PEAI anhydrous powder in a 1:2 molar ratio. They were then dissolved in $\gamma$-butyrolactone so that the perovskite concentration was 0.25 M. The dissolution was carried out for one hour at 50 $^{\circ}$C. After the perovskite was removed from the glovebox, 2 $\mu$l of PEPI-F solution was used for crystallization and placed between the DBR and oxygen plasma-activated glass. The solution thus sandwiched was sealed in a Teflon jar with 2 milliliters of dichloromethane which served as an antisolvent. After 12 hours, the resulting crystals were dried in a nitrogen flow and the thickness was adjusted by mechanical exfoliation using a layer of PDMS as a tape.

\subsection*{AFM measurments}
\begin{figure}
\centering
    \includegraphics[width=0.8\linewidth]{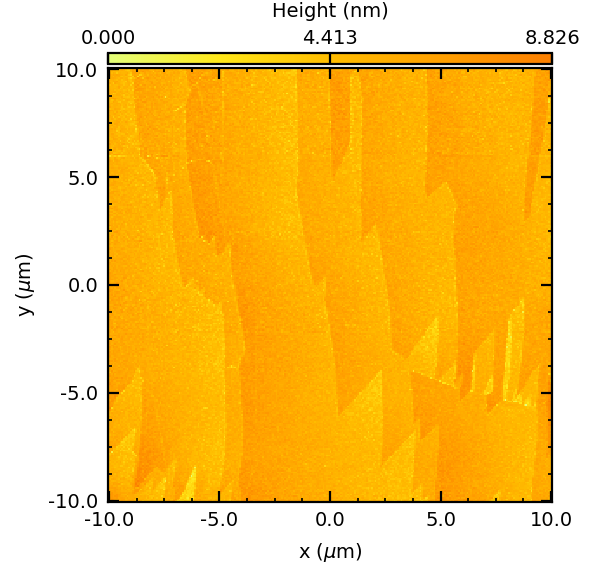}
    \par
    \includegraphics[width=0.8\linewidth]{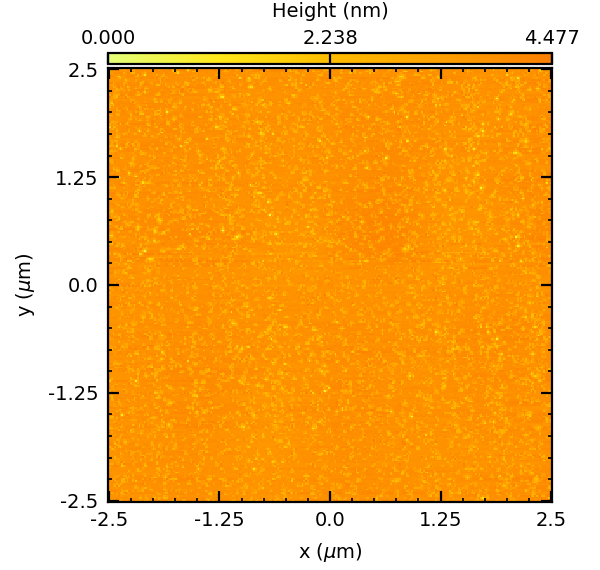}
\caption{Atomic force microscope measurements of the test sample. The upper panel shows a fragment with a large number of faults caused by the layered structure of perovskite, the lower panel shows a homogeneous fragment of the crystal surface.}
  \label{fig:AFM}
\end{figure}

To determine the surface roughness, AFM measurements were taken of the crystal under study. Figure~\ref{fig:AFM} shows a fragment of the crystal with multiple terraces resulting from the layered structure of PEPI-F (top panel) and a fragment of a homogeneous crystal (bottom panel). The calculated RMS of the roughness is 0.46nm

\subsection*{Linewidths}
\begin{figure*}[t]
\begin{center}
\includegraphics[width=0.8\textwidth]{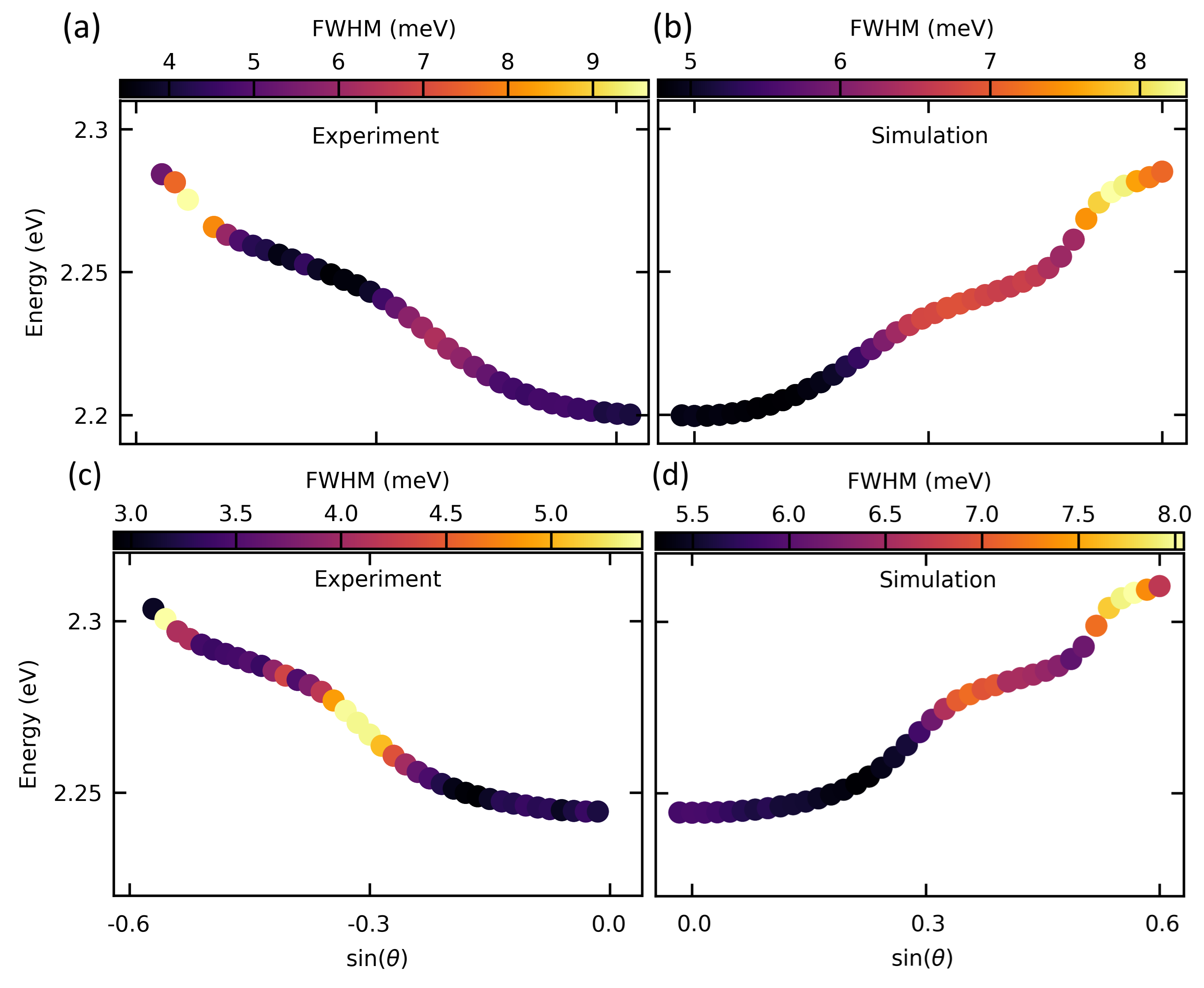}
\caption{Comparison of FWHM for experimental results (a,c) and simulations with the transfer matrix method (b,d) from data illustrated in Figure 3a. Panels (a,b) show the mode having a minimum at an energy of about 2.26 eV  and (c,d) of about 2.20 eV.}
\label{fig:fwhm}
\end{center}
\end{figure*}

In Figure~\ref{fig:fwhm}, we compare experimental full-width at half maximum (FWHM) with values calculated numerically by the transfer matrix method for two polariton modes of different energies. The results obtained exhibit a reasonably good quantitative and qualitative agreement. Here, the narrowest lines occur near the minimum of the mode, where the polariton-photon branch becomes mostly photonic, as opposed to the broadening of the line observed at the inflection points—indicative of the transition to a more polariton-like state.

\subsection*{Air gap simulation}
\begin{figure*}[t]
\centering
\includegraphics[width=\textwidth]{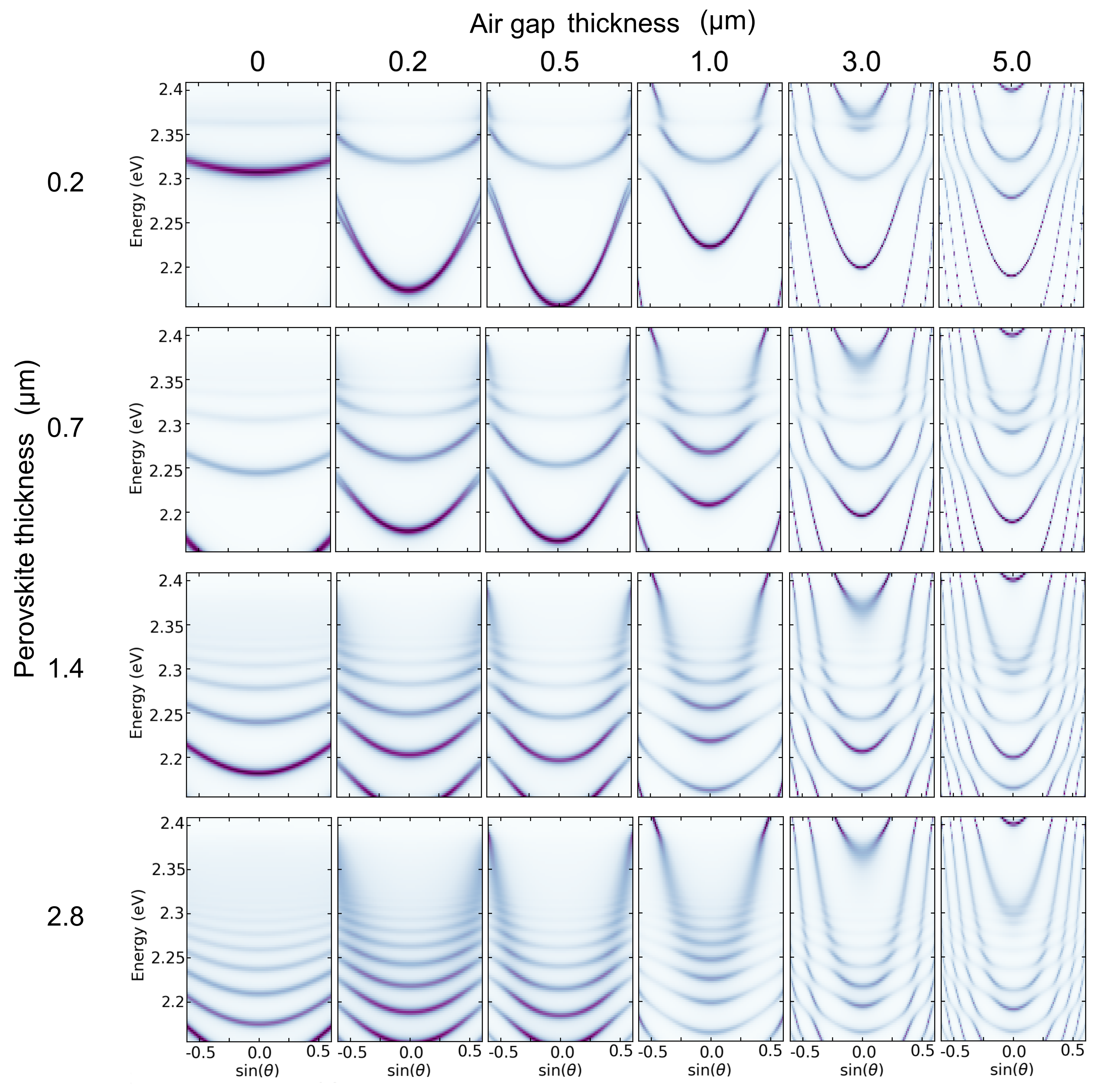}
\caption{ Simulation of the effect of adding an air layer whose thickness in $\mu$m is in the row at the top affects the PEPI-F spectrum of the crystal thickness specified in the column on the left. }
\label{fig:thickness}
\end{figure*}

\begin{figure*}[t]
\centering
\includegraphics[width=\textwidth]{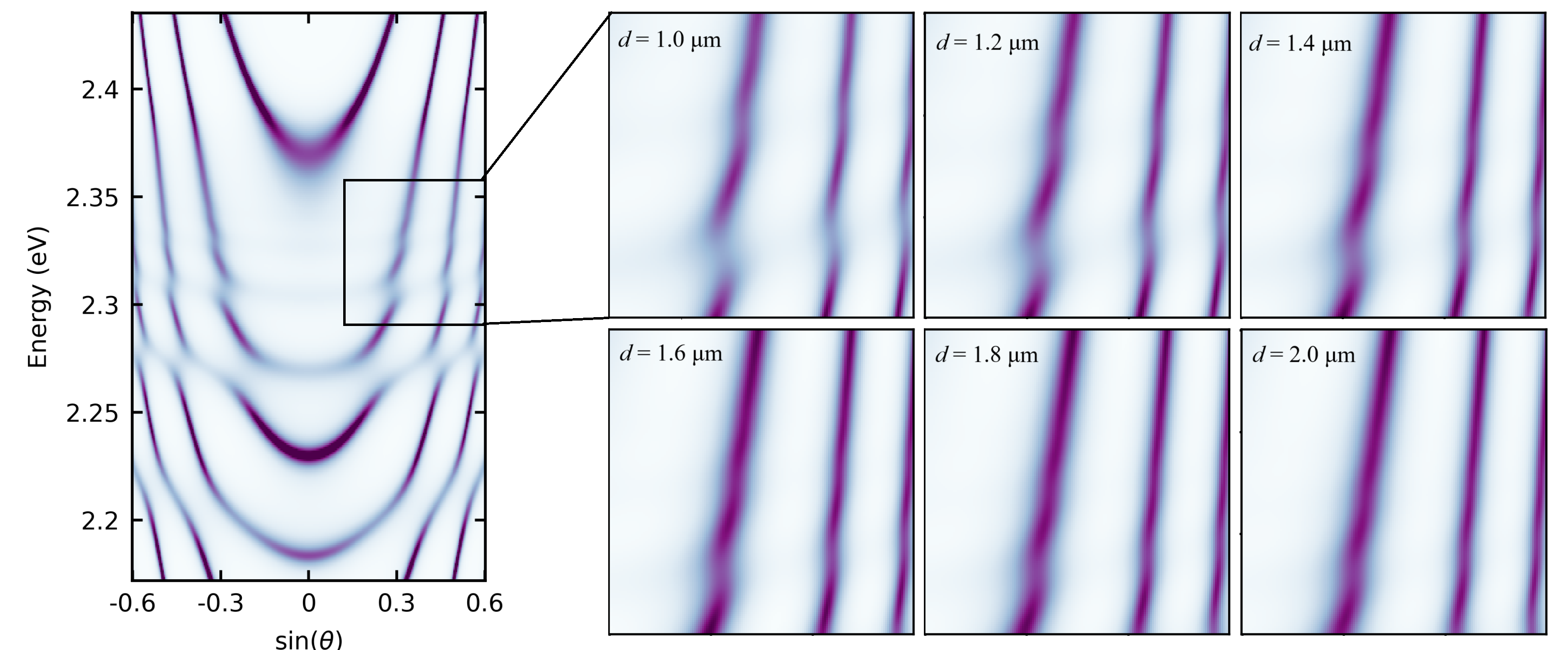}
\caption{ Representation of closing the energy gap by changing the perovskite thickness in the transfer matrix method. }
\label{fig:gap}
\end{figure*}

A simulation of how the addition of an air gap affects the dispersion of a perovskite microcavity can be seen in Figure~\ref{fig:thickness} and \ref{fig:gap}. 

\subsection*{Fitting and simulation parameters}
We performed the fitting dispersion parameters based on the reflection spectrum using the non-linear least squares method from the SciPy package. The fitting parameters are: $E_{C,0}^{(1)}=2328.44$ meV, $E_{C,0}^{(2)}=2440.46$ meV, $E_{C,0}^{(3)}=2546.79$ meV, $E_{C,0}^{(4)}=2653.65$ meV, $E_{C,0}^{(5)}=2764.89$ meV. The rest of the parameters are taken as below: $E_{X,0}=2389.74$ meV, $\hbar\Omega_R=329.10$ meV, $\hbar \gamma_C= 0.00658$ meV, $\hbar\gamma_X= 30$ meV, $\hbar\gamma_E= 0.00658$ meV, $m^*_{C}=0.0016$, $m^*_{E,1}=0.18\cdot10^{-3}$, $m^*_{E,2}=0.23\cdot10^{-3}$,
$m^{*}_{E,3}=0.23\cdot10^{-3}$.
{$E_{E,0}$ for  $CE_1$, $CE_2$ and $CE_3$ are $1997.30$ meV, $2209.60$ meV and $2360.50$ meV.
Parameters $E_{E,0}$ for  $CE_I$ and $CE_{II}$  are  $2032.08$ meV and $2300.0$ meV.} Here, we assume a small change in the external photon effective mass for different modes to achieve the best agreement with the experimental data. The coupling parameters for photonic fraction are: $J_C^{(1,1)}=14.49$ meV,
$J_C^{(1,2)}=9.45$ meV,
$J_C^{(1,3)}=8.50$ meV,
$J_C^{(1,4)}=4.79$ meV,
$J_C^{(1,5)}=3.78$ meV. Coupling parameters for excitonic fraction are given by: 
$J_X^{(1,1)}=48.20$ meV,
$J_X^{(1,2)}=31.50$ meV,
$J_X^{(1,3)}=28.35$ meV,
$J_X^{(1,4)}=15.96$ meV,
$J_X^{(1,4)}=12.60$ meV. We assume that $J_C^{(n,1)}$ equals $J_C^{(n,2)}$ and $J_X^{(n,1)}$ equals $J_X^{(n,2)}$ to reduce the number of free parameters.
The most accurate correspondence between the model and the calculated eigenenergies is achieved by adjusting the $J$ parameters for different $n$ and $l$ values separately. This phenomenon can be attributed to the reduction in strong coupling for higher modes energies. Wavevectors were normalized by taking $k_0$ = 11.86 $\mu m^{-1}$.
For planar microcavities in-plane wavevector is directly related to incidence/reflectance angle $\theta$ through the relation $k_\parallel=\frac{E}{\hbar c} \sin \theta$.

\end{document}